\documentclass[superscriptaddress,floatfix,prl,aps,showpacs]{revtex4}
\usepackage{graphicx}
\usepackage{latexsym}
\usepackage{amsmath}
\begin{document}
\bibliographystyle{apsrev}
\title{A Luttinger Liquid in a Box}
\author{Fabrizio Anfuso}
\affiliation{Institute of Theoretical Physics,
Chalmers University of Technology and G\"oteborg University,
S-412 96 G\"oteborg, Sweden}
\affiliation{Physics Dept., University of Bologna, INFM and INFN,
I-40127 Bologna, Italy}
\author{Sebastian Eggert}
\affiliation{Institute of Theoretical Physics,
Chalmers University of Technology and G\"oteborg University,
S-412 96 G\"oteborg, Sweden}
\date{\today}
\begin{abstract}
We study a Luttinger Liquid in a finite one-dimensional wire with box-like
boundary conditions by considering
the local distribution of the single particle spectral weight.
This corresponds to the experimental probability of extracting a 
single electron at a given place and energy, which can be interpreted as the
square of an electron  wave-function.   For the non-interacting case, this 
is given by a standing wave at the Fermi wave-vector. In the presence
of interactions, however, the wave-functions obtain additional structure with 
a sharp depletion near the edges and modulations throughout 
the wire.  In the spinful case, these modulations correspond to the separate 
spin- and charge-like excitations in the system.
\end{abstract}
\maketitle
The problem of a particle in a one-dimensional box is a classic example 
in almost any quantum mechanics text-book since it gives a pedagogical 
introduction to the concept of energy quantization 
and provides a complete visualization of the corresponding wave-functions.   
It is however only recently that this problem has gained true experimental 
relevance due to the progress in constructing smaller and more refined 
structures to confine electrons in dots and wires.  It is for example 
now possible to resolve the electron wave functions in a finite piece of 
carbon nanotube by Scanning Tunneling Microscopy (STM) 
experiments \cite{venema,lemay,nanotube}.  Most experimental realizations of 
one-dimensional electron boxes contain many electrons in a Fermi sea, but it 
is possible to study a single particle excitation on top of such a ground 
state configuration and classify the possible energy levels 
and wave-functions.  

However, electron-electron interactions may produce interesting effects in
such one-dimen\-sional many-body systems and
systematically change the shape of the electron wave-functions. 
In fact single particle excitations are 
no longer the eigenstates of an interacting Hamiltonian,
but of course it is still interesting to determine the probability of 
extracting or inserting individual electrons at a 
given position and energy. This local 
probability density can be interpreted as the
square of the electron wave-function.
We therefore study the fundamental problem of single particle excitations
in a many body {\it interacting} Fermion system confined to a one-dimensional 
box using the Luttinger Liquid formalism.  

We find that the classic
example of a box provides again a good visualization 
of the effect of interactions on the wave-function and the energy quantization. 
In particular, in addition to the expected rapid 
Fermi wave-vector oscillations in the wave-function we can 
recognize long wavelength modulations, which correspond to the 
underlying boson-like excitations in the Luttinger Liquid.  
For repulsive interactions the wave-functions are
sharply depleted at the edges with a characteristic power-law.
Analytic expressions for the wave-functions of the first few levels are
presented.

The Luttinger Liquid formalism is a well-established tool to describe 
interacting electrons confined to one dimension \cite{haldane,senechal}.
In a linearized region around the Fermi points the Fermion field can 
be expanded in terms of left- and right-movers
\begin{equation}
\Psi(x,t)\approx e^{i k_F x}\psi_R(x,t)+e^{-i k_F x}\psi_L(x,t).
\end{equation}
The Fourier modes of the left- and right-moving Fermion density 
are then represented by bosonic creation and annihilation 
operators, which effectively 
act by ``shifting''  Fermions $m$ steps up or down the 
spectrum.  In the presence of 
interactions it is then possible to solve the model by a Bogoliubov
transformation which mixes the left- and  right-moving bosons.
This transformation can be described by a single ``Luttinger
Liquid parameter'' $g$, which gives the hyperbolic
cosine $\alpha$ and sine $\beta$ of the Bogoliubov transformation 
\begin{equation}
\alpha= \frac{1}{2}\left(\frac{1}{g}+g \right), \ \ \ \ \ 
\beta= \frac{1}{2}\left(\frac{1}{g}-g\right). 
\label{cs}
\end{equation}
Here $g<1$ for repulsive backscattering 
interactions and $g=1$ for no interactions.  Forward scattering does not 
affect this parameter, but rescales the effective Fermi-velocity.

Let us now consider spinless Fermions in a one-dimensional box of length $L$
with fixed boundary conditions $\Psi(0)  = \Psi(L) = 0$.
After bosonization the Fermion fields become exponentials of the boson 
operators in a linearized region around the Fermi points 
\begin{eqnarray}
\psi_R(x,t) & = & \frac{1}{\sqrt{2\pi a}} e^{i\sqrt{4 \pi}(\alpha\phi_R(x,t)
-\beta \phi_L(x,t))} \\
\psi_L(x,t) & = & \frac{1}{\sqrt{2\pi a}} e^{-i\sqrt{4 \pi}(\alpha \phi_L(x,t)-\beta \phi_R(x,t))},
\end{eqnarray}
where $a$ is a short-distance cutoff parameter.
The fixed boundary condition therefore relates the left- and right-moving 
boson fields $\phi_{R}(x,t) = -\phi_L(-x,t)+\phi_0$  and determines 
the mode expansion in terms of
ordinary boson creation and annihilation operators and zero 
modes\cite{gogolin,mattsson}
\begin{equation}
\phi_L(x,t)  = \frac{\phi_0+\tilde{\phi_0}}{2}+Q\frac{x+vt}{2L} 
+\sum_{m=1}^\infty \frac{1}{\sqrt{4\pi m}}
\left(e^{-i\frac{m\pi}{L}(x+vt)}a_m^{}
+e^{i\frac{m\pi}{L}(x+vt)}a_m^\dagger\right)
\label{modes}
\end{equation}
where $[\tilde \phi_0, Q] = i$ and 
$\phi_0 = \frac{\sqrt{\pi}}{2g}$. The eigenvalues of the 
zero modes $Q =(n -n_0-\frac{1}{2}) \sqrt{\pi}/g$ are 
quantized with the number of electrons $n$, where 
$n_0=\frac{k_FL}{\pi}$ corresponds to the number of electrons in the 
ground state.
  
The Hamiltonian is given by 
$H = \frac{\pi v}{L}\sum_m  m a_m^\dagger a_m +\frac{v}{2L}Q^2$,
where $v$ is the renormalized Fermi velocity.
We see that the last term in $H$ resembles a ``charging'' energy  
proportional to the square of the excess number of Fermions,  
but this will not affect our calculations since we always
consider single particle excitations with exactly one additional Fermion
$n = n_0 + 1$.   In general there may also be an additional capacitative 
energy with a corresponding single particle charging energy $E_0$.

The boson excitations become highly degenerate with increasing energy levels which 
are always quantized $\omega_m = m \frac{\pi v}{L}$.
However, we are interested 
in the corresponding {\it Fermion} wave-function of a single
particle excitation on the ground state $\langle \omega_m|x\rangle = 
\langle \omega_m|\Psi^\dagger(x)|0\rangle$. 
 The probability density 
$\rho(\omega_m, x)$ is given by the sum of the corresponding 
degenerate wave-functions $\langle \omega_m|x\rangle$ squared
\begin{equation}
\rho(\omega_m, x) \equiv \sum_\lambda
\left|\langle \omega_m, \lambda| \Psi^{\dagger}(x) |0\rangle\right|^2
\label{rho}
\end{equation}
This is the local density of states which gives the 
experimental probability of tunneling an electron into the system
at energy $\omega_m$ and position $x$. This spectral density can be readily 
evaluated for an equally-spaced spectrum by the 
Fourier transformation of the Fermion Green's function
\begin{equation}
\rho(\omega_m, x)=\frac{v}{2L}\int_0^{2 L/v}\! dt \, e^{i\omega_m t}
\langle\Psi (x,t)\Psi^{\dagger}(x,0)\rangle.
\label{fourier}
\end{equation}
\newline
After defining a ``mixed wave'' 
$\chi_m(x)=\alpha e^{im\frac{\pi x}{L}}+\beta e^{-im\frac{\pi x}{L}}$, 
we find
\newline
\begin{eqnarray}
\langle \psi_L^{}(x,t) \psi^\dagger_L(x,0)\rangle&  = & 
 c \left(\sin{\frac{\pi x}{L}} \right)^{2\alpha\beta}
\exp{\left[\sum^{\infty}_{k=1}\frac{e^{-ik\frac{\pi vt}{L}}}{k}|\chi_k(x)|^2\right]} \label{leftleft}\\
\langle \psi_R^{}(x,t) \psi^\dagger_L(x,0)\rangle&  = & 
-c \left(\sin{\frac{\pi x}{L}} \right)^{2\alpha\beta}
\exp{
\left[\sum_{k=1}^{\infty}\frac{e^{-ik\frac{\pi vt}{L}}}{k}\chi^2_k(x) \right]}.
\label{leftright}
\end{eqnarray}
Here $c = \frac{4^{2 \alpha \beta} }{L}
\left(\frac{2 \pi a}{L}\right)^{2 \beta^2}$ is a non-universal 
cutoff-dependent renormalization parameter which sets the
units in our calculations and suppresses the spectral weight in 
the presence of interactions.
These correlation functions can be simplified to powerlaws of 
sine-functions \cite{gogolin,mattsson,boundary} but the 
form above avoids any singularities
in the integral of Eq.~(\ref{fourier}) since for a given 
level $\omega_m$ we can truncate
the sum in the exponential by $k\leq m$.
For the first few levels we are even able to evaluate $\rho$ analytically
\begin{eqnarray}
\rho(\omega_1, x)&=& c
\left(\sin{\frac{\pi x}{L}} \right)^{2\alpha\beta}
 2{\rm Im}^2\left[\chi_1(x)e^{ik_F x}\right] \nonumber \\
\rho(\omega_2, x)&=& c
\left(\sin{\frac{\pi x}{L}} \right)^{2\alpha\beta}
\left(
{\rm Im}^2\left[\chi^2_1(x)e^{ik_F x}\right]
+{\rm Im}^2\left[\chi_2(x)e^{ik_F x}\right]\right) \label{levels}\\
\rho(\omega_3, x)&=& c
\left(\sin{\frac{\pi x}{L}} \right)^{2\alpha\beta}
\left(
\frac{1}{3} {\rm Im}^2\left[\chi^3_1(x)e^{ik_F x}\right]
+{\rm Im}^2\left[\chi_1(x)\chi_2(x)e^{ik_F x}\right]
+ \frac{2}{3}{\rm Im}^2\left[\chi_3(x)e^{ik_F x}\right]\right)\label{level3}
\nonumber 
\end{eqnarray}
\begin{figure}
\begin{center}\includegraphics[width=.75\textwidth]{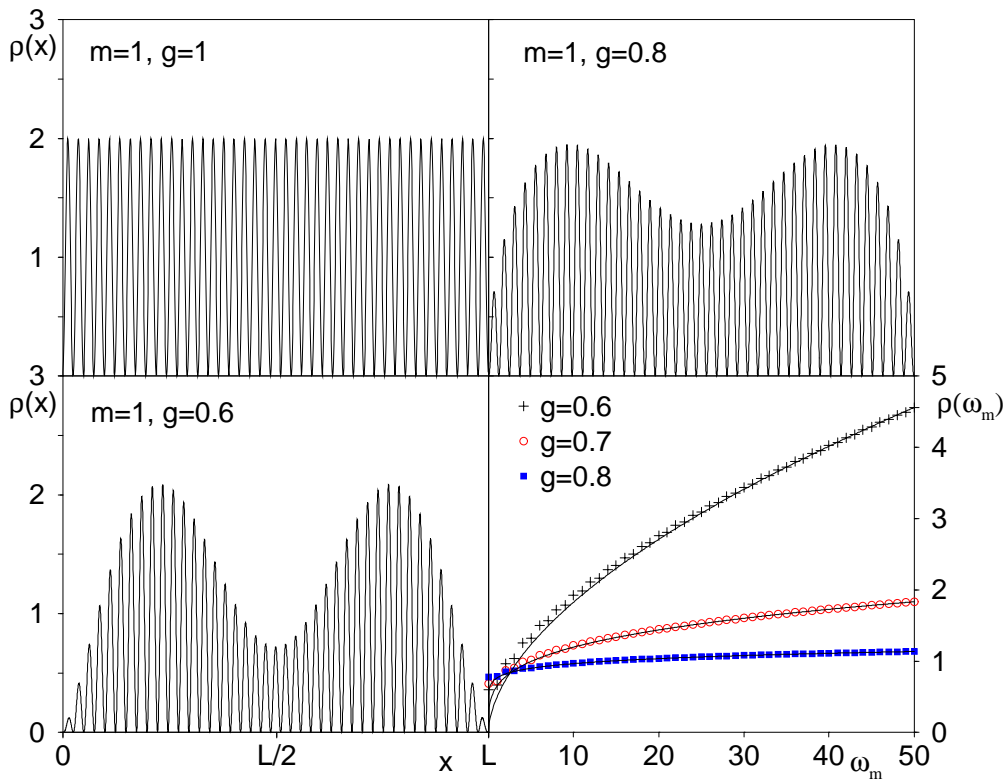}\end{center}
\caption{The wave-function squared corresponding to $m=1$ and $n_0=40$ 
in units of $c$. The modulations become more pronounced with increasing 
interaction strength (smaller g). Last panel:  the integrated
probability density for the first 50 levels compared to the powerlaw behavior
$\omega^{2 \beta^2}$.}
\label{spinless}
\end{figure}
The probability density 
shows an oscillation of $2 k_F x$ with modulations according 
to the mixing of left- and right-moving components in $\chi_m$.  
As can be seen in Fig.~\ref{spinless}, the amplitude of the 
modulations increases
with the interaction strength (smaller $g$)
and the envelope shows a depletion near the 
edges with a characteristic powerlaw $x^{2 \alpha \beta}$ in agreement with the 
notion of a boundary exponent \cite{gogolin,mattsson,boundary,kane}.  
In the limit of small level spacing $L\to \infty$ we recover the known 
Luttinger Liquid powerlaw behavior of the integrated density of states
$\rho(\omega) \propto \omega^{2\beta^2}$ (last panel of Fig.~\ref{spinless}).
In Fig.~\ref{nodes} we see that the modulations of level $\omega_m$
always have $m$ ``nodes'' and  $m+1$ maxima 
with roughly equal spacing and height resembling a 
standing wave with a small wave-vector $\omega_m/v$,
corresponding to the density waves from the
boson excitations relative to the Fermi-energy.
It is also instructive to consider the non-interacting limit $g=1$, for which
we always recover a normalized
standing wave of wave-vector $2(k_F + \omega_m/v)$ without any modulations or depletion, 
even though a general single Fermion excitation still corresponds
to a superposition of degenerate many-boson states (and vice versa). 
\begin{figure}
\begin{center}\includegraphics[width=.75\textwidth]{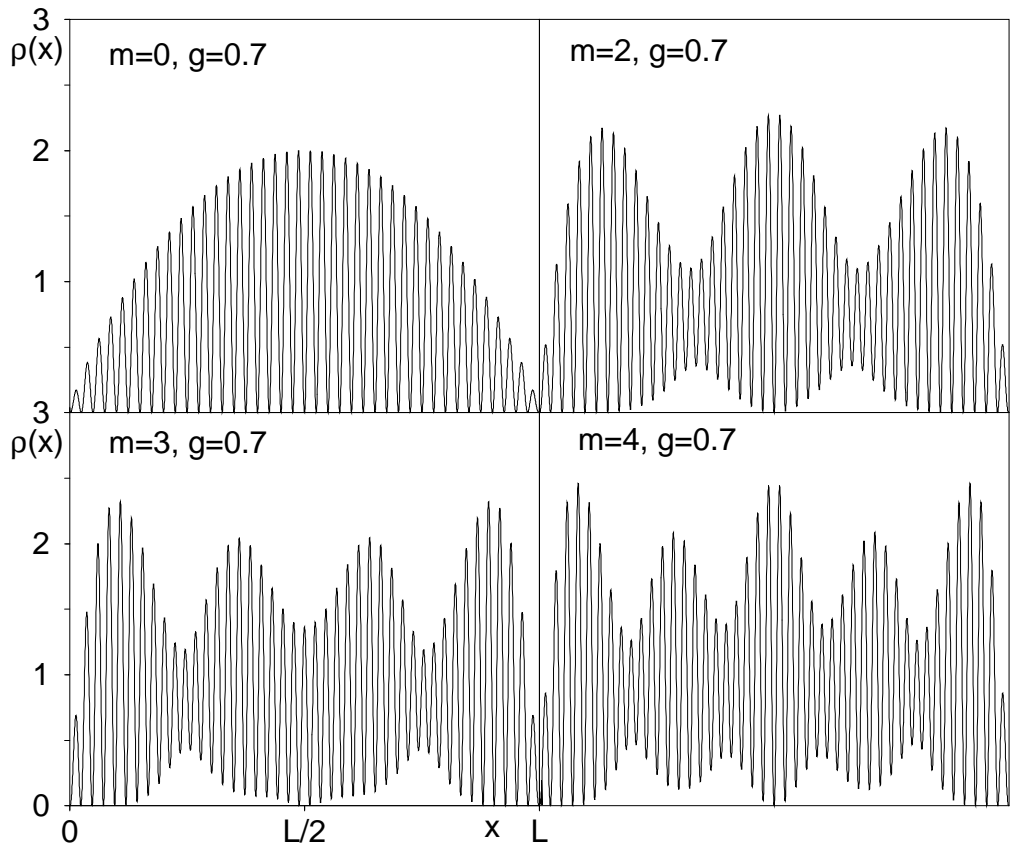}\end{center}
\caption{The wave-function squared of the first few levels for $g=0.7$ and $n_0=40$ in 
units of $c$.}
\label{nodes}
\end{figure}

It is now interesting to explore this modulation pattern  for the spinful
case where we expect separate spin- and charge-like excitations.  In this case
the electron field with spin $\sigma=\pm$ is expressed in terms of spin and 
charge boson operators with two Luttinger Liquid parameters $g_s$ and $g_c$
\begin{equation}
\psi_{R,\sigma}(x) = \frac{1}{\sqrt{2\pi a}} 
e^{i\sqrt{2 \pi} (\alpha_{c}\phi_{R,c} -\beta_{c} \phi_{L,c})}
e^{i\sigma \sqrt{2 \pi} (\alpha_{s}\phi_{R,s} -\beta_{s} \phi_{L,s})}
\end{equation}
and the analogous expression for $\psi_{L,\sigma}$.  The mode expansions for the
spin and charge bosons are the same as in Eq.~(\ref{modes}), and the Hamiltonian is
also given by the a simple sum $H=\sum_{\nu=c,s} \left(\sum_{m_\nu} 
\frac{v_{\nu}\pi }{L} m_\nu a^\dagger_{m_\nu} a^{}_{m_\nu} +\frac{v_{\nu}Q_{\nu}^2}{2L}\right)$.
\begin{figure}
\begin{center}\includegraphics[width=.68\textwidth]{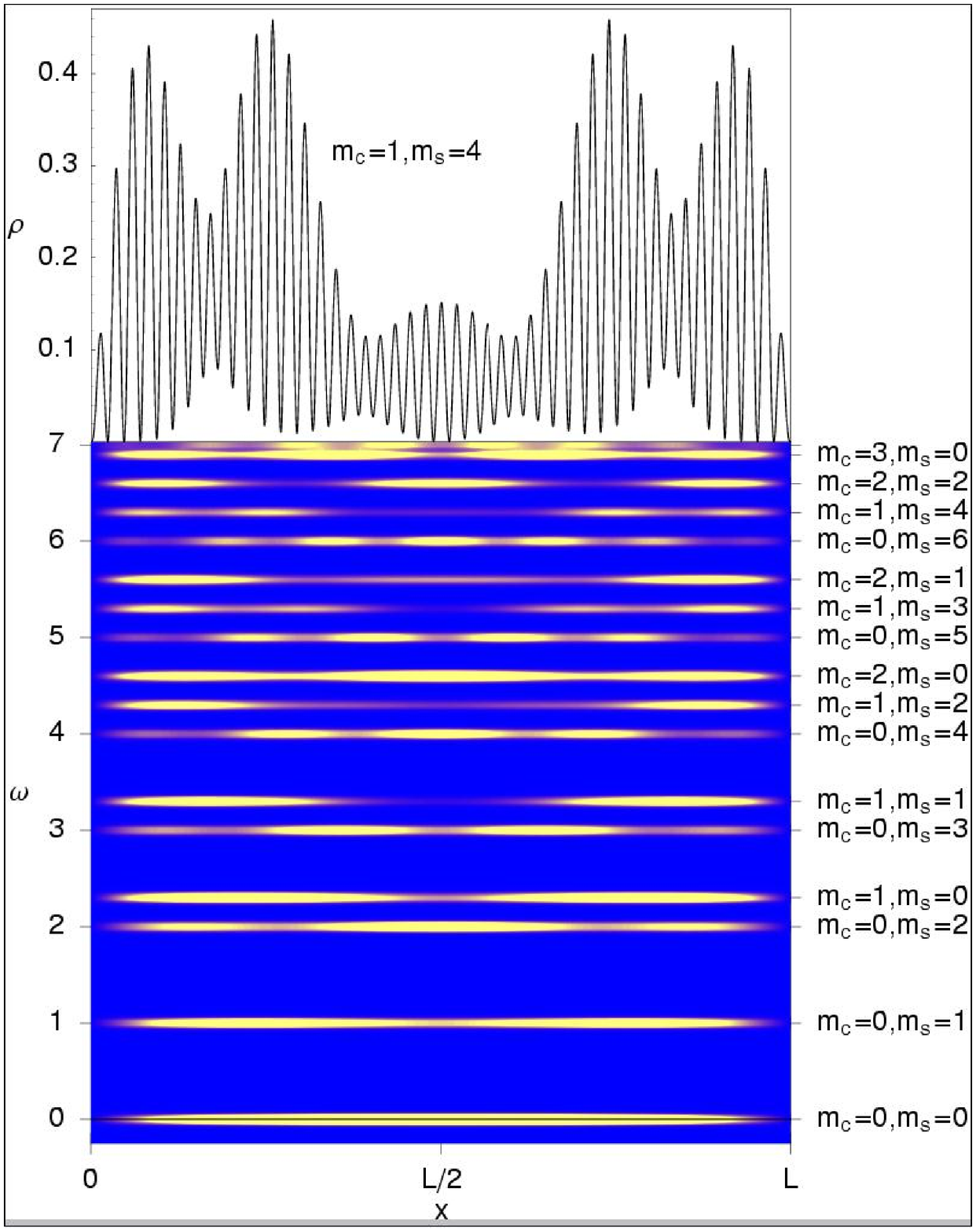}\end{center}
\caption{The single particle density of a spinful interacting 
electron system as a function of energy and position along a  
finite box for $g_s=0.7, \ g_c=0.6$, and $v_c=2.3v_s$ (slightly broadened, 
so that the fast oscillations are averaged out). This may resemble the outcome 
of a $dI/dV$ measurement in an STM experiment on a finite interacting wire.
Top: fully resolved level for $m_s = 4, \ m_c=1$, and $\frac{k_F L}{\pi}=40$.
The two charge peaks and the five weaker spin peaks are visible.}
\label{stm}
\end{figure}
Therefore, the spin and charge excitations are decoupled
except for the quantization conditions on the zero modes 
$Q_s=\frac{\sqrt{\pi}}{\sqrt{2}g_s}l$,  
$Q_c=\frac{\sqrt{\pi}}{\sqrt{2}g_c}\left(n-1-\frac{2 k_F L}{\pi}\right)$ 
that $[l,n]$ must be
either both even or both odd ([1,1] in our case).  
For the non-interacting case, spin- and charge-excitations 
are exactly degenerate, but now the excitations are classified 
by the product space
of two evenly spaced boson spectra with different energy spacing $v_s< v_c$.  
The partition function and the electron Green´s function factorize.
Therefore, the wave-functions are products of spin- and charge-modulations 
which are similar to the ones shown in Fig.~\ref{nodes} and Eq.~(\ref{levels}), 
except that the mixed wave is rescaled
$\chi_{m,\nu}(x) = 
\frac{\alpha_\nu}{\sqrt{2}} e^{im\frac{\pi x}{L}}+\frac{\beta_\nu}{\sqrt{2}} 
e^{-im\frac{\pi x}{L}}, \ \nu=s,c$ and the overall factor is given by
 $\left(\sin{\frac{\pi x}{L}} \right)^{\alpha_s\beta_s}
 \left(\sin{\frac{\pi x}{L}} \right)^{\alpha_c\beta_c}$.
 
In Fig.~\ref{stm} we show the square of the wave-functions  at the 
lowest energies in a spinful 
interacting electron system with $g_s=0.7, \ g_c=0.6$, and $v_c=2.3v_s$.
Due to the different velocities, the degeneracy is lifted and many more
levels appear as the energy is increased [see also Fig.~4 in 
Ref.~\cite{mattsson}].  Each level is classified by a spin and a charge 
quantum number $m_s$ and $m_c$, which is reflected by   
the corresponding number of nodes and maxima in the wave-functions.
For example the level $m_s=4,\ m_c=1$ shows a superposition of two 
charge maxima and five weaker spin maxima.
In Fig.~\ref{stm} we have chosen values of $g_s=0.7$ and $g_c=0.6$, which   
emphasizes the locations of both spin and charge modulations, 
but in systems where
the interactions are mostly spin-independent $g_s$ is likely to 
be much closer to unity and the spin modulations are less pronounced.  
Nonetheless, the level structure due to
$v_s<v_c$ as well as the charge modulations are likely to be observable 
when classifying Luttinger Liquid systems in real space with STM 
experiments.

The integrated weight of the individual levels 
decreases with increasing energy,
but the total (averaged) density of states increases with
the known powerlaw $\rho(\omega) \propto \omega^{\beta_s^2+\beta_c^2}$.
The superstructures survive even in the continuum limit
and give rise to the observed slow oscillations with wave-vectors $\omega/v_c$ and
$\omega/v_s$ near the edge of 
a semi-infinite Luttinger Liquid \cite{mattsson,boundary,stm,meden}.
In the non-interacting limit $g_s=g_c=1$ we recover again 
standing waves with a fixed wave-vector without any modulations, and 
remarkably all the many degenerate spin and charge boson states at each
energy level exactly 
sum up to an integrated normalized spectral weight of unity.  

 In conclusion we have shown that the single particle wave-functions 
of interacting Fermions in a box can be used to visualize the true 
nature of the underlying boson excitations. Given the rapid progress
in STM imaging and nano-structured materials this may soon be observable when
classifying different kinds of potential Luttinger Liquid systems.

This research was supported in part by the Swedish Research Council and INFM.


\begin{thebibliography}{999}
\bibitem{nanotube}  Due to the 
metallic substrates, interaction effects are screened in the two 
experiments \cite{venema,lemay}.
\end{thebibliography}
\end{document}